# Kondo physics in carbon nanotubes

Jesper Nygård[*], David Henry Cobden[†], & Poul Erik Lindelof[*]

[*]Ørsted Laboratory, Niels Bohr Institute, Universitetsparken 5, DK-2100 Copenhagen, Denmark
[†]Department of Physics, University of Warwick, Coventry CV4 7AL, UK

The connection of electrical leads to wire-like molecules is a logical step in the development of molecular electronics, but also allows studies of fundamental physics. For example, metallic carbon nanotubes[1] are quantum wires that have been found to act as one-dimensional quantum dots[2,3], Luttinger-liquids[4,5], proximity-induced superconductors[6,7] and ballistic[8] and diffusive[9] one-dimensional metals. Here we report that electrically-contacted single-wall nanotubes can serve as powerful probes of Kondo physics, demonstrating the universality of the Kondo effect. Arising in the prototypical case from the interaction between a localized impurity magnetic moment and delocalized electrons in a metallic host, the Kondo effect has been used to explain[10] enhanced low-temperature scattering from magnetic impurities in metals, and also occurs in transport through semiconductor quantum dots[11-18]. The far higher tunability of dots (in our case, nanotubes) compared with atomic impurities renders new classes of Kondo-like effects[19,20] accessible. Our nanotube devices differ from previous systems in which Kondo effects have been observed, in that they are one-dimensional quantum dots with three-dimensional metal (gold) reservoirs. This allows us to observe Kondo resonances for very large electron number ($N$) in the dot, and approaching the unitary limit (where the transmission reaches its maximum possible value). Moreover, we detect a previously unobserved Kondo effect, occurring for even values of $N$ in a magnetic field.

Each of our devices[21] contains a metallic single-walled nanotube with gold source and drain contacts and a substrate gate contact, as illustrated in Fig. 1a. Several measures are taken to optimize electrical contact between the gold and the nanotubes (see Methods). The resulting two-terminal linear-response conductance $G$ ranges up to more than $3e^2/h$ (resistance 8 kΩ) at room temperature, where $h$ is Planck's constant and $e$ is the electronic charge. We believe this is the highest reported yet for a single-walled tube[22]. This implies that the transmission probability $P$ of each contact can reach about 0.9 (the maximum theoretical conductance of a metallic tube is[1] $4e^2/h$), demonstrating that nearly ideal contacts are achievable between a normal metal and a molecule. Meanwhile, the variation of $P$ from device to device allows the investigation of different transport regimes.

For $P \ll 1$, as is well established[2,3], Coulomb blockade occurs at low temperature $T$ and the device behaves as a one-dimensional (1D) quantum dot[23]. The values of the charging energy, $U$ (the average interaction between two electrons), and the average level spacing, $\Delta E$, found for this type of device are as expected for a tube segment of effective length $L$ equal to the contact spacing[3]. At higher $T$, the device exhibits a power-law variation of $G$ with $T$, which can be well explained by Luttinger-liquid behaviour[4]. At the other extreme, for $P \approx 0.9$, we see no Coulomb blockade, and instead the characteristics[21] resemble those of a diffusive 1D wire even for $T < 1$ K.

It is the intermediate transmission regime which we focus on here. The measurements in Fig. 1 were made on a device having $G \approx 1.6\ e^2/h$ (we estimate $P \sim 0.6$) at room temperature. At low $T$, $G$ undergoes large, reproducible fluctuations as a function of substrate gate voltage $V_g$ (Fig. 1b). Fig. 1c shows the variation with $T$ over a narrower range of $V_g$. At $T = 780$ mK, the fluctuations are regular and can be identified as Coulomb blockade oscillations. However, unlike normal Coulomb blockade, as $T$ is decreased $G$ decreases in some valleys, marked above with an E, while it increases in other valleys, marked with an O. Fig. 1d shows a greyscale plot of the differential conductance d$I$/d$V$ vs source-drain bias $V$ and $V_g$ at $T \sim 75$ mK. The enhanced conductance in the O valleys corresponds to the dark (high d$I$/d$V$) horizontal features at $V = 0$, such as those labelled X and Y. No such lines are seen in the E valleys, which are instead bounded by roughly hexagonal bubbles. Although the distinctness of these features varies with $V_g$, O- and E-type valleys can be seen to alternate for as many as ten consecutive peaks.

These features can be explained by invoking the Kondo effect[11-13]. When the number $N$ of electrons on the dot is odd, the total spin $S$ is a half-integer, and second-order processes such as that shown on the left of Fig. 1f can occur. Similar processes, which change the total spin of the dot, add up coherently to form a correlated many-electron state in which electrons in the two leads are strongly coupled, allowing current flow even under blockade conditions if $T < T_K$, the Kondo temperature. When $N$ is even, however, there is no equivalent process, and thus no current, if $S = 0$. This matches the data if in the O valleys $N$ is odd and $S = ½$ while in the E valleys $N$ is even and $S = 0$. Further evidence confirming that features X and Y in Fig. 1d are "Kondo ridges" is provided by the logarithmic dependence[16] of $G$ on $T$ in the centers of the ridges, shown in Fig. 1e, and by the splitting of the ridges in a magnetic field (see later).

This new Kondo system differs in several important ways from the two-dimensional (2D) semiconductor dots in which Kondo physics was previously studied.[14-19] First, for semiconductor dots the leads are 2D electron systems, while for tube dots they are normal 3D metal (gold). Second, the excitation spectrum of an interacting 1D dot, which behaves as a Luttinger liquid when Coulomb blockade is overcome[4], is profoundly different from a 2D dot. Third, since semiconductor dots are electrostatically defined, their geometry and contact transmission probabilities depend on gate voltages. In contrast, for tube dots the contacts cannot be individually tuned, but the fixed geometry means that signs of the Kondo effect can be seen over the entire range of $V_g$, encompassing hundreds of Coulomb oscillations. Fourth, the effect of a magnetic field is entirely through spin in single-walled nanotube dots, while orbital effects are often dominant in semiconductor dots. Fifth, thanks to this fact combined with the absence of spin-orbit coupling, unlike for a semiconductor dot the spin of a tube dot is well defined and easily measured[24,25] even for large $N$. Sixth and last, the conditions[14,23] on the size $L$ for observing the Kondo effect are weaker in the tube dot. Halfway along a Kondo ridge, $T_K \sim (\Gamma U)^{1/2} \exp[-(\pi/4)U/\Gamma]$, where $\Gamma$ is the level width. The ratio $U/\Gamma$ cannot be made smaller than $U/\Delta E$, because $\Gamma < \Delta E$ for any quantum dot. For 2D semiconductor dots, $U/\Delta E \propto L$, so that the maximum possible $T_K$ decreases exponentially with $L$. This requires them to be made as small as possible, so limiting the electron number $N$ to less than ~ 100. In contrast, for tube dots, which are 1D, $U/\Delta E \sim 6$ is independent[3] of length $L$, and the maximum $T_K$ decreases only with the prefactor $(\Gamma U)^{1/2}$. This helps to explain how tube dots can show such a strong and clear Kondo effect, even for large $N$. An order of magnitude estimate of $N$ is the number of π-electrons in the tube, ie, the number of carbon atoms, which is tens of thousands. (Without knowing the exact dot size or band structure a more accurate estimate is not possible.) Nanotube quantum dots therefore allow the first ever observations of the Kondo effect in the limit of very large $N$.

The width $\Gamma$ varies from level to level, so we can search for pairs of peaks with particularly large $\Gamma$, such as those illustrated in Figs. 2a and 2b. For both these pairs the valley between the peaks is completely gone at base temperature (~75 mK). In the Kondo regime, $G$ is expected to saturate at the unitary limit $G_0$ as $T \to 0$, and at higher $T$ it should follow a universal scaling form[10] which can be approximated by[15] $G(T) = G_0/(1+(2^{1/s}-1)(T/T_K)^2)^s$, where $s = 0.22$ for spin ½ on the dot. To test this, we fit $G$ vs $T$ in the valley center to this expression using $T_K$ and $G_0$ as fitting parameters. The results, shown in Fig. 2c, are highly satisfactory. For both peak pairs we find $G_0 > 1.7e^2/h$, which is close to the maximum value of $2e^2/h$ obtained for symmetric coupling to the two contacts (this assumes the double Fermi-point degeneracy[1] is broken). The deduced values of $T_K$ of 1.6 K and 0.9 K are in good agreement with nonlinear measurements, as indicated in the insets above Figs. 2a and 2b. Here, on top of greyscale plots of $dI/dV$ showing the Kondo ridges are superimposed traces of $dI/dV$ vs $V$ midway along each ridge. The Kondo effect is expected to be suppressed[13] on a bias scale $|V| \sim k_B T_K/e$. Indeed, we see that the width of each peak in $dI/dV$ is roughly $2k_B T_K/e$ (indicated by the vertical white bars). We believe these first measurements on a nanotube dot are closer to the unitary limit than has ever been reported in a semiconductor quantum dot.

In some regions of $V_g$, such as that shown in Fig. 3, a regular series of faint diamonds can be discerned. This resembles the standard behaviour of a quantum dot for $P \ll 1$, where only first-order tunnelling is significant and Coulomb blockade forces $N$ to be fixed at an integer within the diamonds. However, superimposed on the diamonds are a variety of horizontal features which can be attributed to higher-order tunnelling processes that are strong in this device because $P$ is large. Kondo ridges are seen at $V = 0$ in the odd-$N$ diamonds (marked O again), which are outlined by white dashed lines. Additional horizontal ridges are seen at $V \neq 0$, such as those labelled P in the even and Q in the odd diamonds. Features of type P truncate every even diamond, resulting in the characteristic alternating bubble-ridge pattern seen also in Fig. 1d. It is natural to associate a horizontal feature at a bias $V = \Delta/e$ with processes generating an excitation of energy $\Delta$ in the dot. An example of such a process is cotunneling involving two nondegenerate single-particle states, as sketched on the right of Fig. 3. However, since such inelastic cotunneling normally only produces a step in $dI/dV$ above a threshold, we suggest that these peaks in $dI/dV$ at finite bias offsets are another manifestation of Kondo physics.

Finally, we look at the effect of magnetic field $B$. Fig. 4a shows the evolution with $B$ of an adjacent bubble (E) and Kondo ridge (O). As expected,[13] the Kondo ridge splits linearly into components at $V = \pm g\mu_B B/e$ (the Zeeman energy), where $\mu_B$ is the Bohr magneton and $g = 2.0$ is the electron g-factor[2,25]. Meanwhile, the edges of the bubble move linearly towards $V = 0$. Fig. 4b shows the $dI/dV$ characteristics in the center of the bubble. At a certain field, $B_c = 1.18$ T, the bubble collapses to a single ridge, corresponding to a peak in $dI/dV$ at $V = 0$. Its $T$ dependence is shown in Fig. 4c. The peak height is logarithmic in $T$ (Fig. 4d), just as for the odd-$N$ Kondo ridge at $B = 0$. A similar ridge is formed in most bubbles at a field dependent on the bubble width at $B = 0$.

Exactly such behaviour has recently been predicted[20] for quantum dots when $N$ is even, in the case where the Zeeman energy is sufficiently large. It is assumed that at $B = 0$ the ground state is a singlet ($S = 0$), and the lowest excited state is a triplet ($S = 1$) at a distance $\Delta_{t-s}$ above the ground state. In an applied field $B_c = \Delta_{t-s}/g\mu_B$, the $S_z = -1$ member of the triplet becomes degenerate with the singlet (see Fig. 4e). At this point, $S$ alternates between 1 and 0 as electrons cotunnel between the contacts (see Fig. 4f), resulting in a new type of Kondo resonance[20] with similar characteristics to the conventional type seen at $B = 0$. This theory is particularly appropriate to single-walled nanotubes, where the effects of $B$ are almost entirely through spin. Applying the model, we deduce for instance that the edges of the E bubble in Fig. 4a are associated with a singlet-triplet excitation of energy $\Delta_{t-s} = g\mu_B B_c = 137$ µeV.

Note that this new type of Kondo effect is distinct from the one recently discovered in a semiconductor dot[19], which occurs when a singlet is aligned with a *degenerate* triplet by orbital effects in a weak magnetic field. It serves to underline the fact that new physical phenomena can emerge in the study of molecular nanostructures.

**Methods**

In making devices, the as-grown single-walled nanotube material[26] is sonicated in dichloroethane and deposited on the $SiO_2$ by diluting with isopropyl alcohol before drying. To improve the contacts: the leads are fabricated as soon as possible after tube deposition by low-energy (10 kV) electron-beam lithography; the polymethylmethacrylate resist is over-exposed to avoid residue; the contact metallization is pure gold, evaporated directly on top of the tubes; and each contact covers at least 0.5 µm of the tube/bundle. Although we cannot resolve the difference between a single tube and a thin bundle of tubes in the atomic force microscope, we study only devices whose detailed characteristics imply that the conduction is determined by a single metallic nanotube[21].

**Acknowledgements.** We thank A. Rinzler and R. Smalley for supplying the nanotubes, K. G. Rasmussen, M. M. Andreasen, A. E. Hansen and A. Kristensen for experimental assistance, and M. Pustilnik, N. Wingreen, L. P. Kouwenhoven, N. d'Ambrumenil, P. R. Poulsen and P. L. McEuen for helpful discussions.

Correspondence should be addressed to D.H.C. (email: d.h.cobden@warwick.ac.uk).

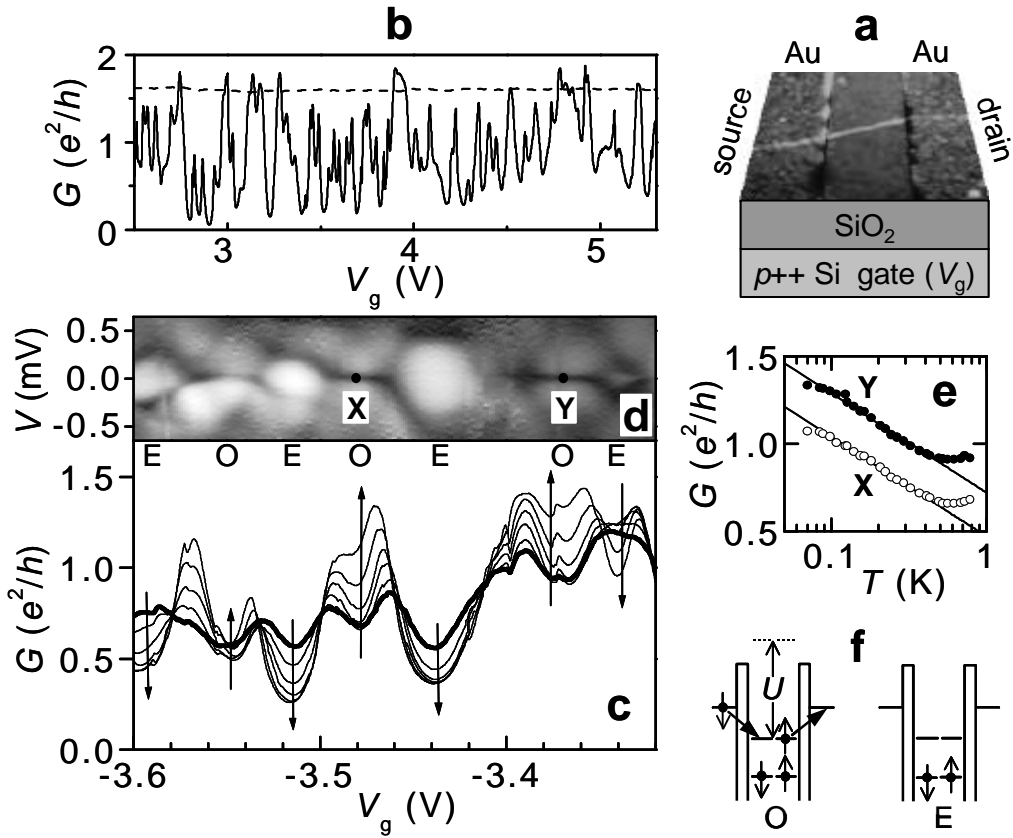

**Figure 1.** Characteristics of a nanotube device with intermediate contact transmission probabilities. **a** Device consisting of a 2 nm-thick nanotube bundle with 30 nm-thick gold contacts separated by 300 nm. Sketched beneath the tapping-mode atomic force microscope (AFM) image are the 300-nm thick $SiO_2$ layer and the highly $p$-doped Si substrate used as the gate electrode. **b** At room temperature (dashed line) the linear-response conductance $G$ is about 1.6 $e^2/h$, almost independent of gate voltage $V_g$. At $T \approx$ 75 mK, the base electron temperature on our dilution refrigerator, reproducible fluctuations are seen in $G$ vs $V_g$. (The source-drain bias $V$ used was 7 µV dc). **c** Temperature dependence of a narrower region of $V_g$. $T$ = 75, 125, 180, 245, 320, 490, 560, and (thicker line) 780 mK. Vertical arrows indicate the direction of change with decreasing $T$, which is opposite in the regions labelled E and O. **d** Greyscale plot of $dI/dV$ against $V$ and $V_g$, over the same region of $V_g$ (darker = more positive). **e** Conductance vs temperature at points X and Y in panel (d). The straight lines indicate $\log(T)$ behaviour. **f** Interpretation of the situation in the O and E regions. In O regions the number of electrons $N$ on the dot is odd, and higher-order spin-flipping processes can occur, leading to a Kondo resonance. In the E regions $N$ is even and no such processes exist.

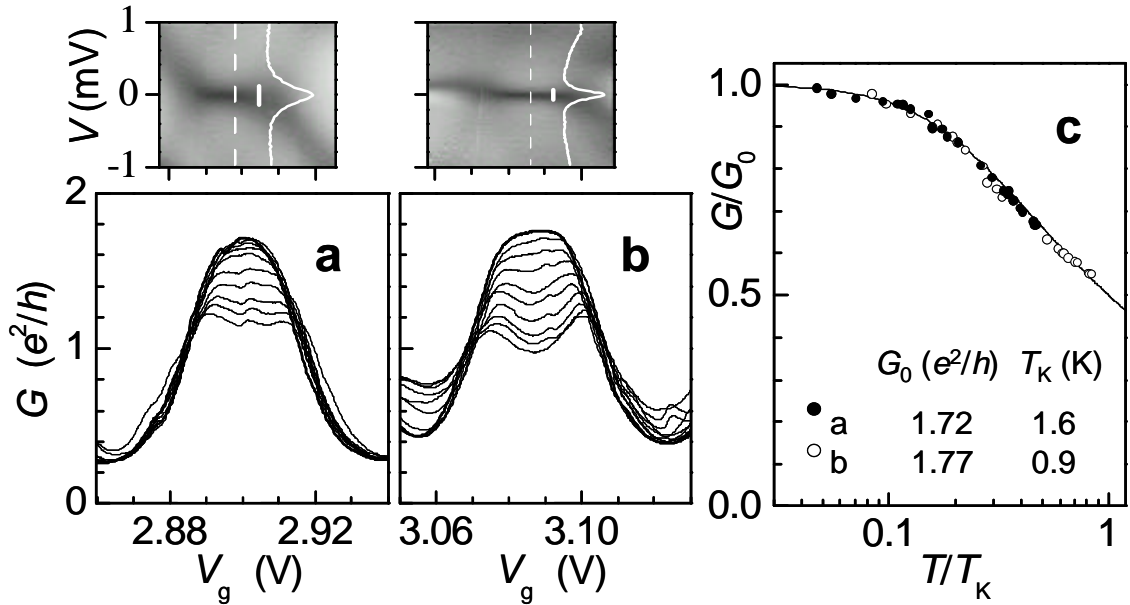

**Figure 2.** Analysis of two high-conductance peak pairs. **a** and **b** Temperature dependence of $G$ vs $V_g$. $T = 75$ (thick line), 87, 150, 200, 250, 330, 420, 530, 630 and 740 mK. Above are $dI/dV$ greyscales at 75mK in the same regions of $V_g$, showing broad Kondo ridges. Superimposed on each is the $dI/dV$ vs $V$ trace along the dashed line, which also serves to denote $dI/dV = 0$. The vertical white bars are of length $2k_B T_K/e$, where $T_K$ is obtained in each case from the data in panel (c). **c** $G$ vs $T$ data in the valley centers for the peak pairs in panels (a) and (b), scaled in each case to match the theoretical scaling function given in the text (solid line). The deduced values of $G_0$ and $T_K$ are given.

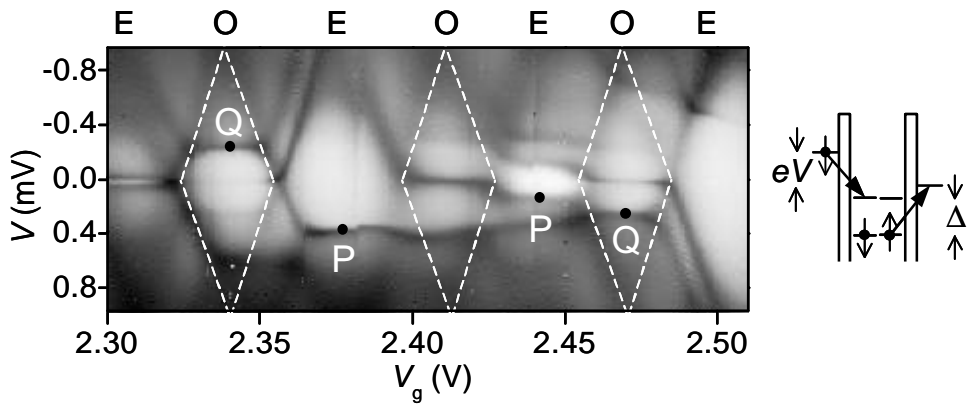

**Figure 3.** $dI/dV$ greyscale plot in a different $V_g$ region at $T = 75$ mK. Dashed lines outline the odd Coulomb diamonds. Horizontal features such as those labelled P and Q suggest higher-order processes like the second-order one sketched on the right, where tunneling involves two levels separated by an energy $\Delta$ and occurs only for bias $V \geq \Delta/e$.

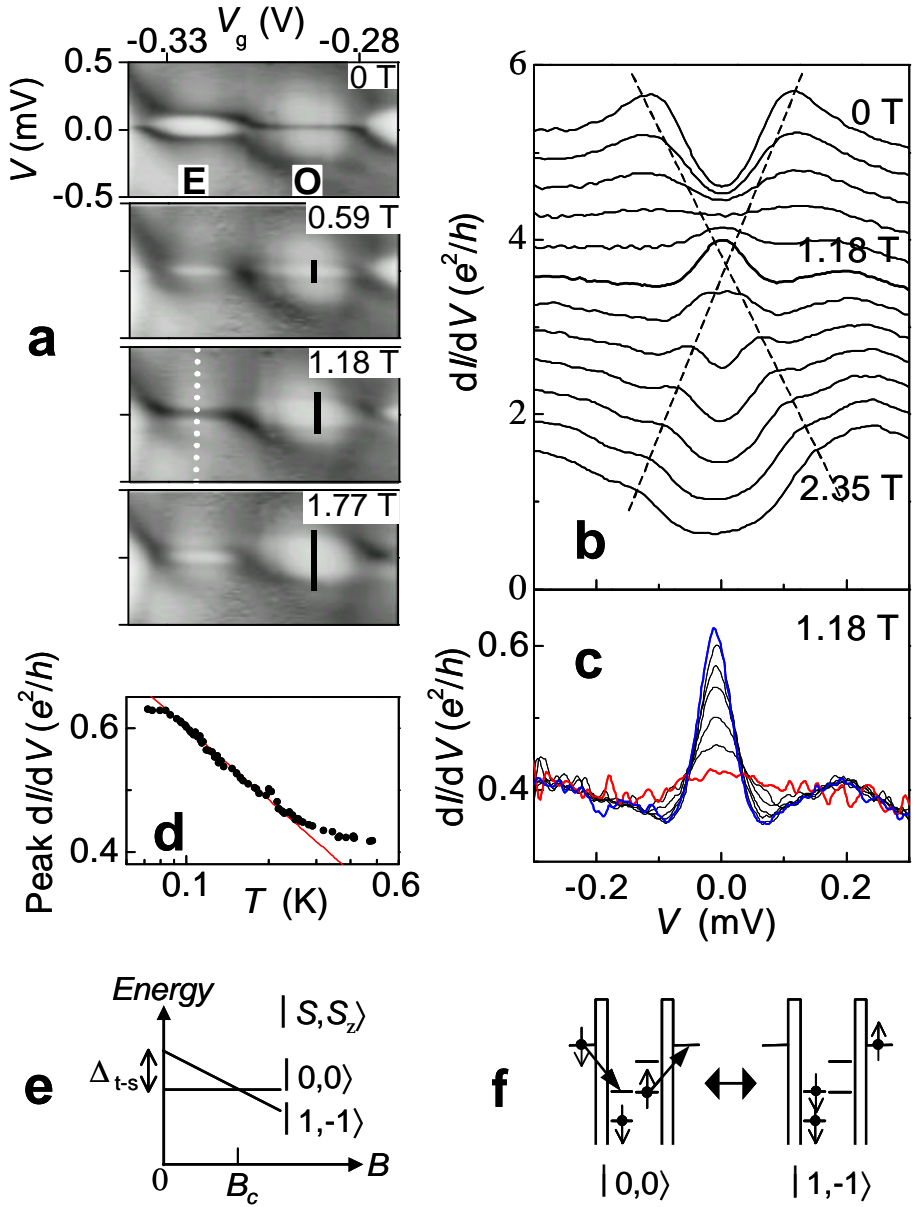

**Figure 4**. Effect of perpendicular magnetic field $B$. **a** $dI/dV$ greyscale plot at $T = 75$ mK showing adjacent odd-$N$ (O) and even-$N$ (E) regions at a series of fields. The vertical bars have length $2g\mu_B B/e$. As $B$ increases the O ridge splits while the E bubble shrinks to a single ridge at 1.18 T before reappearing at higher $B$. **b** Evolution of $dI/dV$-$V$ characteristics at $V_g = -0.322$ V (indicated by the white dotted line in (a)) at $T = 75$ mK. The traces are offset from each other by 0.4 $e^2/h$ for clarity. The dashed lines, which indicate the motion of the peaks, are sloped according to $g\mu_B B/e$. **c** Temperature dependence of the peak in $dI/dV$ at $B = 1.18$ T. $T = 75$ (blue), 100, 115, 130, 180, 230, and 350 mK (red). **d** Peak value of $dI/dV$ (at $V = 0$) plotted against $\log(T)$. **e** A singlet $|S,S_z\rangle = |0,0\rangle$ ground state becomes degenerate with a triplet $|1,-1\rangle$ excited state at $B =$

$B_c = \Delta_{t-s}/g\mu_B$. At this point, spin-flipping higher order transitions, as sketched in **f**, become possible, leading to a new type of Kondo resonance.